\documentclass[showpacs,preprintnumbers,amsmath,amssymb,twocolumn]{revtex4}
\usepackage{graphicx}
\usepackage{dcolumn}
\usepackage{bm}
\usepackage{amssymb}
\usepackage{array}
\usepackage{hhline}
\usepackage{longtable}

\begin{document}

\title{Explicit determination of mean first-passage time for random walks on deterministic uniform recursive trees}

\author{Zhongzhi Zhang$^{1,2}$}
\email{zhangzz@fudan.edu.cn}

\author{Yi Qi$^{1,2}$}

\author{Shuigeng Zhou$^{1,2}$}
\email{sgzhou@fudan.edu.cn}

\author{Shuyang Gao$^{1,2}$}

\author{Jihong Guan$^{3}$}
\email{jhguan@tongji.edu.cn}

\affiliation {$^{1}$School of Computer Science, Fudan University,
Shanghai 200433, China}

\affiliation {$^{2}$Shanghai Key Lab of Intelligent Information
Processing, Fudan University, Shanghai 200433, China}

\affiliation{$^{3}$Department of Computer Science and Technology,
Tongji University, 4800 Cao'an Road, Shanghai 201804, China}

\begin{abstract}
The determination of mean first-passage time (MFPT) for random walks
in networks is a theoretical challenge, and is a topic of
considerable recent interest within the physics community. In this
paper, according to the known connections between MFPT, effective
resistance, and the eigenvalues of graph Laplacian, we first study
analytically the MFPT between all node pairs of a class of growing
treelike networks, which we term deterministic uniform recursive
trees (DURTs), since one of its particular cases is a deterministic
version of the famous uniform recursive tree. The interesting
quantity is determined exactly through the recursive relation of the
Laplacian spectra obtained from the special construction of DURTs.
The analytical result shows that the MFPT between all couples of
nodes in DURTs varies as $N \ln N$ for large networks with node
number $N$. Second, we study trapping on a particular network of
DURTs, focusing on a special case with the immobile trap positioned
at a node having largest degree. We determine exactly the average
trapping time (ATT) that is defined as the average of FPT from all
nodes to the trap. In contrast to the scaling of the MFPT, the
leading behavior of ATT is a linear function of $N$. Interestingly,
we show that the behavior for ATT of the trapping problem is related
to the trapping location, which is in comparison with the phenomenon
of trapping on fractal T-graph although both networks exhibit tree
structure. Finally, we believe that the methods could open the way
to exactly calculate the MFPT and ATT in a wide range of
deterministic media.
\end{abstract}


\date{\today}
\maketitle

\section{Introduction}


The field of complex networks has been very active in the past
decade, since they have been proven a powerful tool to describe very
diverse systems in nature and
society~\cite{AlBa02,DoMe02,Ne03,BoLaMoChHw06,DoGoMe08}. One of the
ultimate goals of the study of complex networks is to understand the
influences of network structure on dynamics running on
them~\cite{Ne03,BoLaMoChHw06,DoGoMe08}. Among various dynamical
processes, random walks on networks are fundamental to many branches
of science and engineering and have received a surge of interest in
recent
years~\cite{NoRi04,NoRi04E,SoRebe05,CoBeMo05,CoBeMo07,CoBeTeVoKl07,GaSoHaMa07,BaCaPa08}.
As a basic dynamical process, random walks are relevant to a variety
of aspects of complex networks, such as target
problem~\cite{JaBl01}, community detection~\cite{ErSiMasn03,NeGi04},
network routing~\cite{PaAm01}, reaction-diffusion
processes~\cite{CoPaVe07,YuKaKi09}, and so on. Therefore, it is of
major theoretical interest and  practical importance to investigate
random walks on complex networks.

A primary quantity of interest that relates to random walks is the
first-passage time (FPT) defined as the expected time for a walker
to first reach a given destination node starting from a source
point. The importance of FPTs originates from the following main
aspects. First, their first encounter properties are relevant to
those in a plethora of real situations~\cite{CoBeTeVoKl07}, such as
transport, disease spreading, target search, to name a few. On the
other hand, they can measure the efficiency of random navigation on
networks~\cite{CaDe09}. Last but not least, many other quantities
for random walks can be expressed in terms of FPTs, and much
information about the random-walks dynamics is encoded in
FPTs~\cite{Re01}. Recently, there have been a growing number of
theoretical studies on
FPTs~\cite{NoRi04,CoBeMo05,CoBeTeVoKl07,BaCaPa08}. 
Many authors have devoted their endeavors to study the average of
FPTs to hit a given target node from all other nodes~\cite{Mo69,
KaBa02PRE,KaBa02IJBC,Ag08,ZhQiZhXiGu09,ZhGuXiQiZh09,ZhXiZhGaGu09,ZhLiGaZhGuLi09}. 
In addition, relevant work also addressed the FPTs between all
couples of nodes, giving some numerical
results~\cite{HuXuWuWa06,CaAb08} or approximate
scalings~\cite{CoBeMo05,CoBeMo07,CoBeTeVoKl07,GaSoHaMa07,Bobe05,ZhZhZhYiGu09,ZhZhXiChGu09}.
These simulation results and scaling laws are necessary as a first
step toward understanding random walks on networks; however they do
not provide a complete picture of the random-walk dynamics, and
analytical exact solutions are helpful in this regard~\cite{Gi96}.

In this paper, we study analytically random walks on a class of
deterministic treelike networks. By using the links between the
random walks, electrical networks, and Laplacian spectra, we first
compute exactly the mean first-passage time (MFPT) between two nodes
over all pairs of nodes. The obtained explicit formula indicates
that for large networks with $N$ nodes, the MFPT is asymptotic to $N
\ln N$. Then, we study the trapping problem, a particular
random-walk issue, on a special case of the network family with a
trap fixed at a node of the highest degree. We derive rigorously the
average tapping time (ATT), which is the average of FPTs from all
nodes to the trap. We show that in contrast to the scaling of MFPT,
the leading behavior of ATT grows lineally with $N$. Since the MFPT
can be considered as the average of ATT with the trap distributed
uniformly on all nodes of the entire network, we conclude that the
trap location has an important influence on the behavior of ATT. We
expect that the our analysis technique could be applicable to
determining MFPT and ATT for a broad range of deterministic
networks.

\section{The deterministic uniform recursive trees}

\begin{figure}
\begin{center}
\includegraphics[width=0.85\linewidth,trim=60 05 60 0]{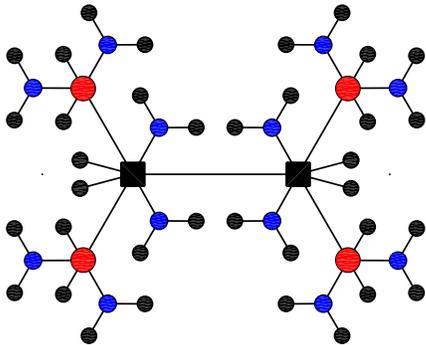}
\caption{(Color online) The first several iterative processes of a
particular network for the case of $m=2$.} \label{network}
\end{center}
\end{figure}

We first introduce the model concerned, which is a class of trees
(networks) defined in an iterative method~\cite{JuKiKa02}. Let
$U_{g}$ ($g\geq 0$) denote the networks after $g$ iterations. Then
the networks can be generated as follows. Initially ($g=0$), $ U
_{0}$ has two nodes connected by an edge. For $g \geq 1$, we can
obtain $U_{g}$ from $U_{g-1}$ by adding $m$ ($m$ is a natural
number) new offspring nodes to each existing node in $U_{g-1}$. At
each iteration $g_i$ ($g_i\geq 1$), the number of newly generated
nodes is $L(g_i) =2\,m\,(m +1)^{ g_i -1}$. Thus, the order (i.e.,
number of nodes) and the number of edges in $U_{g}$ are
$N_g=\sum_{g_i=0}^{g}L(g_i)= 2\,(m +1)^{g}$ and $K_g=N_g-1= 2\,(m
+1)^{g}-1$, respectively. Figure~\ref{network} shows an example of
the network family for a special case of $m=2$ after four
iterations.

Note that the network for the special case of $m=1$ is actually a
deterministic version of the uniform recursive tree
(URT)~\cite{SmMa95}, which is a principal famous
model~\cite{DoKrMeSa08,ZhZhZhGu08} for random graphs~\cite{ErRe60}
and has a variety of important applications in many
aspects~\cite{Mo74,NaHe82,Ga77}. Moreover, since this particular
case of the networks under consideration has similar topological
characteristics as the URT, we call the investigated networks
\emph{deterministic uniform recursive trees} (DURTs), which could be
helpful for better understanding of the nature of the URT.

We study this model because of its intrinsic
interest~\cite{DoMeOl06,BaCoDaFi08,ZhZhQiGu08,QiZhDiZhGu09,ZhQiZhLiGu09}
and its relevance to real-life networks. For instance, it is
small-world~\cite{JuKiKa02,DoMeOl06,ZhZhQiGu08,QiZhDiZhGu09,WaSt98};
particularly, the so-called border tree motifs have been shown to be
present, in a significant way, in real-world
systems~\cite{ViroTrCo08}. In the rest of this paper, we will study
random walks performed on DURTs with an aim to better understand
dynamical process occurring on them.

\section{\label{sec03}Formulating standard random-walks on DURTs}

We study a simple model for random walks on the DURTs $U_{g}$. At
each time step, the walker makes a jump from its current location to
any of its nearest neighbors with uniform
probability. 
We are interested in the FPT of a random walker starting from a
source to a given target point, averaged over all node pairs of
source and target points.


To determine the FPT between a pair of two different nodes, one can
make use of the method of the pseudoinverse of the Laplacian
matrix~\cite{BeGr03} for $U_{g}$, where random walks are performed.
It is a very efficient method, which allows to obtain the FPT
between two arbitrarily distinct nodes directly from the network
topology and only requires inversion of a single $N_{g} \times
N_{g}$ matrix. The pseudoinverse of the Laplacian matrix,
$\textbf{L}_g^\dagger$, of $U_{g}$, is in fact a variant of the
inverse of its Laplacian matrix, $\textbf{L}_g$. The elements
$l_{ij}^{g}$ of the latter are defined as follows: $l_{ij}^{g}=-1$
if nodes $i$ and $j$ are connected by a link, otherwise
$l_{ij}^{g}=0$; while $l_{ii}^{(g)}=k_i$ (viz., degree of node $i$).
Then, the pseudoinverse of the Laplacian matrix $\textbf{L}_g$ is
defined to be~\cite{RaMi71}
\begin{equation}\label{Pinverse01}
 \textbf{L}_g^\dagger=\left(\textbf{L}_g-\frac{\textbf{e}_g\textbf{e}_g^\top}{N_g}\right)^{-1}+\frac{\textbf{e}_g\textbf{e}_g^\top}{N_g}\,,
\end{equation}
where $\textbf{e}_g$ is the $N_g$-dimensional ``one" vector, i.e.,
$\textbf{e}_g=(1,1,\ldots,1)^\top$.

We use $F_{ij}(g)$ to denote the FPT for the walker in $U_{g}$,
starting from node $i$ to node $j$, which can be expressed in terms
of the entries $l_{ij}^{\dagger,g}$ of $\textbf{L}_g^\dagger$ as
follows~\cite{CaAb08}: 
\begin{equation}\label{Hitting01}
 F_{ij}(g)=\sum_{n=1}^{N_g}\left(l_{in}^{\dagger,g}-l_{ij}^{\dagger,g}-l_{jn}^{\dagger,g}+l_{jj}^{\dagger,g}\right)l_{nn}^{g}\,,
\end{equation}
where $l_{nn}^{g}$ is the $n$ entry of the diagonal of the Laplacian
matrix $\textbf{L}_g$. So the total, $F_{\rm tot}(g)$, for FPTs
between all pairs of nodes in $U_{g}$ reads
\begin{equation}\label{Hitting02}
 F_{\rm tot}(g)=\sum_{i\neq j}\sum_{j=1}^{N_g}F_{ij}(g)\,,
\end{equation}
and the MFPT averaged over all node pairs, $\langle F \rangle_g$, is
then
\begin{equation}\label{Hitting03}
 \langle F \rangle_g=\frac{F_{\rm tot}(g)}{N_g(N_g-1)}=\frac{1}{N_g(N_g-1)}\sum_{i\neq
 j}\sum_{j=1}^{N_g}F_{ij}(g)\,.
\end{equation}

\begin{figure}
\begin{center}
\includegraphics[width=0.80\linewidth,trim=30 30 20 0]{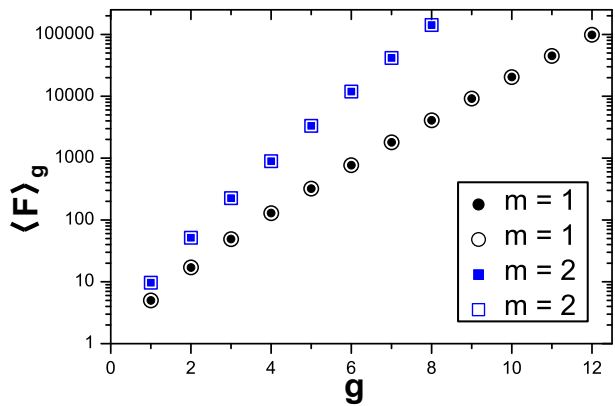}
\end{center}
\caption[kurzform]{\label{Time01} (Color online) Mean first-passage
time $\langle F \rangle_g$ as a function of the iteration $g$ on a
semilogarithmic scale for different values of parameter $m$. The
filled symbols are the numerical results obtained by direct
calculation from Eqs.~(\ref{Hitting01}) and (\ref{Hitting03}), while
the empty symbols correspond to the analytical values from
Eq.~(\ref{Hitting150}), both of which are consistent with each
other.}
\end{figure}

Equations~(\ref{Hitting01}) and (\ref{Hitting03}) show that the
issue of computing $\langle F \rangle_g$ is reduced to finding the
elements of the pseudoinverse matrix $\textbf{L}_g^\dagger$. 
Since for large $g$ the network order $N_g$ increases exponentially
with $g$, it becomes intractable to obtain $\langle F \rangle_g$
through direct calculation using the pseudoinverse matrix, because
of the limitations of time and computer memory, and one can compute
directly the MFPT only for the first iterations (see
Fig.~\ref{Time01}). 
Thus, it would be satisfactory if good methods could be proposed to
get around this problem. Fortunately, the particular construction of
the DURTs and the connection~\cite{ChRaRuSm89,Te91} between
effective resistance and the FPTs for random walks allow us to
calculate analytically the MFPT to obtain a rigorous solution.
Details will be provided below.

\section{Exact solution to MFPT averaged over all node pairs}

In order to avoid the computational complexity of inverting the
matrix, in what follows, we will use the connection between the
electronic networks and random walks to find a closed-form formula
for MFPT, $\langle F \rangle_g$.

\subsection{Relation for commute time and effective resistance between two nodes}

For a given graph $G$, its underlying electrical
network~\cite{DoSn84} can be obtained by replacing each edge of $G$
with a unit resistor. 
The effective resistance of electrical network provides an
alternative way to compute FPTs for random walks on the original
network~\cite{ChRaRuSm89,Te91}. It has been proven that for a
connected graph, the FPTs, $F_{ij}$ and $F_{ji}$ between nodes $i$
and $j$, and the effective resistance, $R_{ij}$, between these two
nodes satisfy the following relation:
\begin{equation}\label{Hitting04}
F_{ij}+F_{ji}=2\,K\,R_{ij}\,,
\end{equation}
where $K$ is the number of all edges in the graph and $F_{ij}$ is
the expected time that a random walker spends on reaching node $j$
for the first time, starting from node $i$. Actually, the sum,
$F_{ij}+F_{ji}$, is the average time for a walker to go from $i$ to
$j$ and back or vice versa, and it is often called commute
time~\cite{GoJa74} represented by $C_{ij}$, i.e.,
$C_{ij}=F_{ij}+F_{ji}$. By symmetry, $C_{ij}=C_{ji}$. Then,
Eq.~(\ref{Hitting04}) can be recast as
\begin{equation}\label{Hitting05}
C_{ij}=C_{ji}=2\,K\,R_{ij}\,.
\end{equation}

Thus, if we view $U_{g}$ as resistor networks by considering each
edge to be unit resistor, according to the close relation between
FPTs and effective resistance shown in Eq.~(\ref{Hitting04}) and
(\ref{Hitting05}), Eq.~(\ref{Hitting02}) can be rewritten as
\begin{equation}\label{Hitting06}
 F_{\rm tot}(g)=\frac{1}{2}\sum_{i\neq j}\sum_{j=1}^{N_g}C_{ij}(g)=K_{g}\,\sum_{i\neq
 j}\sum_{j=1}^{N_g}R_{ij}(g)\,,
\end{equation}
where $C_{ij}(g)$ and $R_{ij}(g)$ represent respectively the commute
time and effective resistance between two nodes $i$ and $j$ of
$U_{g}$. Analogously, Eq.~(\ref{Hitting03}) can be recast in terms
of effective resistances as
\begin{equation}\label{Hitting07}
 \langle F \rangle_g=\frac{F_{\rm tot}(g)}{N_g(N_g-1)}=\frac{1}{N_g}\sum_{i\neq
 j}\sum_{j=1}^{N_g}R_{ij}(g)\,,
\end{equation}
where the sum of effective resistors between all pairs of nodes of
$U_{g}$ is the so-called Kirchhoff index~\cite{BoBaLiKl94}, which we
denote by $R_{\rm tot}(g)$. Using the previously obtained
results~\cite{GuMo96,ZhKlLu96}, the following relation holds:
\begin{equation}\label{Hitting08}
R_{\rm tot}(g)=\sum_{i\neq
 j}\sum_{j=1}^{N_g}R_{ij}(g)=2\,N_g\,\sum_{i=2}^{N_g}\frac{1}{\lambda_i}\,,
\end{equation}
where $\lambda_i$ ($i=2,\ldots, N_g$) are all the nonzero
eigenvalues of Laplacian matrix, $\textbf{L}_g$, of network $U_{g}$.
Note that since $U_{g}$ is connected, its Laplacian matrix has only
one zero eigenvalue $\lambda_1$, i.e., $\lambda_1=0$. Then, we have
\begin{equation}\label{Hitting09}
 \langle F \rangle_g=2\,\sum_{i=2}^{N_g}\frac{1}{\lambda_i}\,.
\end{equation}
Having $\langle F \rangle_g$ in terms of the sum of the reciprocal
of all nonzero Laplacian eigenvalues, the next step is to determine
this sum.

\subsection{Determining MFPT using Laplacian eigenvalues}

After reducing the problem to finding the total of the reciprocal of
all nonzero eigenvalues of $\textbf{L}_g$, in the following text, we
will resolve this problem.

By construction, it is easy to derive the following recursion
relation between $\mathbf{L}_g$ and $\mathbf{L}_{g-1}$,
\begin{eqnarray}\label{matrix03}
\mathbf{L}_g =\left(\begin{array}{ccccc}\textbf{L}_{g-1}+m\textbf{I}
& \textbf{-I}& \textbf{-I} &\cdots& \textbf{-I}
\\\textbf{-I} &{\textbf{I}}& {\textbf{0}}&\cdots& {\textbf{0}}
\\\textbf{-I} & {\textbf{0}}& {\textbf{I}}&\cdots& {\textbf{0}}
\\\vdots & \vdots & \vdots & \  &\vdots
\\\textbf{-I} & {\textbf{0}}& {\textbf{0}}&\cdots& {\textbf{I}}
\end{array}\right),
\end{eqnarray}
where each block is a $2(m+1)^{t-1}\times 2(m+1)^{t-1}$ matrix and
$\textbf{I}$ is the identity matrix. Then, using the elementary
matrix operations and the results in~\cite{Si00}, the characteristic
polynomial, $P_g(x)$, of $\textbf{L}_g$
satisfies~\cite{ZhQiZhLiGu09}

\begin{eqnarray}\label{matrix04}
P_g(x)&=&{\rm det}(x\textbf{I}-\textbf{L}_g)\nonumber\\
&=&\big\{{\rm det}[(x-1)\textbf{I}]\big\}^m \cdot {\rm
det}\left(\left(x-m-\frac{m}{x-1}\right)\textbf{I}-\textbf{L}_{g-1}\right),
\end{eqnarray}
which can be rewritten recursively as
\begin{equation}\label{matrix05}
P_g(x)=(x-1)^{2m(m+1)^{g-1}}\times P_{g-1}(\varphi(x)),
\end{equation}
where
\begin{equation}\label{matrix06}
\varphi(x)=x-m-\frac{m}{x-1}\,.
\end{equation}

Since there are $2(m+1)^g$ nodes in $U_g$, the Laplacian matrix
$\textbf{L}_g$ has $2(m+1)^g$ eigenvalues, which are represented as
$\lambda^g_1, \lambda^g_2,\ldots,\lambda^g_{2(m+1)^g}$,
respectively. We denote by $E_g$ the set of these Laplacian
eigenvalues, i.e., $E_g=\{\lambda^g_1,
\lambda^g_2,\ldots,\lambda^g_{2(m+1)^g}\}$, and we assume that
$\lambda^g_1 \le \lambda^g_2 \le \ldots \le \lambda^g_{2(m+1)^g}$.
The set $E_g$ can be classified into two subsets represented by
$E_g^{(1)}$ and $E_g^{(2)}$, respectively~\cite{ZhQiZhLiGu09}. That
is to say, $E_g=E_g^{(1)} \cup E_g^{(2)}$, where $E_g^{(1)}$
consists of eigenvalue 1 with multiplicity $2(m-1)(m+1)^{g-1}$,
\begin{equation}\label{subset01}
E_g^{(1)}=\{\underbrace{1,1,1,\dots,1,1}_{2(m-1)(m+1)^{g-1}\mbox{}}\}\,,
\end{equation}
in which the distinctness of elements has been ignored.

The remaining $4(m+1)^{g-1}$ eigenvalues of $\textbf{L}_g$, forming
the subset $E_g^{(2)}$, are determined by equation
$P_{g-1}(\varphi(x))=0$ and expressed separately by
$\bar{\lambda}^g_1,
\bar{\lambda}^g_2,\ldots,\bar{\lambda}^g_{4(m+1)^{g-1}}$. For the
sake of convenience, we presume $\bar{\lambda}^g_1 \le
\bar{\lambda}^g_2 \le \ldots \le \bar{\lambda}^g_{4(m+1)^{g-1}}$.
Thus, $E_g^{(2)}=\big \{\bar{\lambda}^g_1,
\bar{\lambda}^g_2,\ldots,\bar{\lambda}^g_{4(m+1)^{g-1}}\big \}$.

According to Eq.~(\ref{matrix06}), it is obvious that for an
arbitrary element in $E_{g-1}$, say $\lambda_{i}^{g-1} \in E_{g-1}$,
both solutions of $x-m-\frac{m}{x-1}=\lambda_{i}^{g-1}$ belong to
$E_g^{(2)}$. To facilitate the following computation, we rewrite
equation $x-m-\frac{m}{x-1}=\lambda_{i}^{g-1}$ in an alternative way
as
\begin{equation}\label{matrix07}
x^2-(\lambda_{i}^{g-1}+m+1)x+\lambda_{i}^{g-1}=0\,.
\end{equation}
Moreover, we use notations $\bar{\lambda}_{i}^{g}$ and
$\bar{\lambda}_{i+2(m+1)^{g-1}}^{g}$ to represent it two solutions,
which provide a natural increasing order of the Laplacian
eigenvalues of $U_g$~\cite{ZhQiZhLiGu09}. Solving Eq.
(\ref{matrix07}), its two roots are obtained to be
\begin{equation}\label{matrix08}
\bar{\lambda}_{i}^{g}=\frac{1}{2}\Big(\lambda_{i}^{g-1}+m+1-\sqrt{(\lambda_{i}^{g-1}+m+1)^2-4\lambda_{i}^{g-1}}\Big),
\end{equation}
and
\begin{equation}\label{matrix00}
\bar{\lambda}_{i+2(m+1)^{g-1}}^{g}=\frac{1}{2}\Big(\lambda_{i}^{g-1}+m+1+\sqrt{(\lambda_{i}^{g-1}+m+1)^2-4\lambda_{i}^{g-1}}\Big),
\end{equation}
respectively. Inserting each of the $2(m+1)^{g-1}$ elements of
$E_{g-1}$ into Eqs. (\ref{matrix08}) and~(\ref{matrix00}), one
obtains the subset $E_g^{(2)}$ with cardinality $4(m+1)^{g-1}$.
Considering $E_0=\{0,2\}$ and recursively applying
Eqs.~(\ref{matrix08}) and~(\ref{matrix00}), all Laplacian
eigenvalues of $U_{g}$ can be fully determined.

Having obtaining the recursive solutions of the Laplacian spectra of
$U_{g}$, we continue to calculate the sum on the right-hand side of
Eq.~(\ref{Hitting09}), which is represented by $S_g$ henceforth.
Note that although we fail to determine all the eigenvalues of
$\textbf{L}_g$ in an explicit way, we will show that it is possible
to provide a closed-form expression for $S_g$. By definition, we
have
\begin{eqnarray}\label{Hitting10}
S_g&=& \sum\limits_{i=2}^{2(m+1)^g} \frac{1}{\lambda_i^g} \nonumber\\
&=& \sum_{\lambda_i^g \in
E_g^{(1)}}\frac{1}{\lambda_i^g}+\sum\limits_{i=2}^{4(m+1)^{g-1}}\frac{1}{\bar{\lambda}^g_i}\,.
\end{eqnarray}
we denote the two sums by $S_g^{(1)}$, and $S_g^{(2)}$,
respectively. From Eq.~(\ref{subset01}), we can easily get the first
sum,
\begin{equation}\label{Hitting11}
 S_g^{(1)}=2(m-1)(m+1)^{g-1}\,.
\end{equation}
The second sum can be evaluated as
\begin{eqnarray}\label{Hitting12}
S_g^{(2)}&=& \sum\limits_{i=2}^{4(m+1)^{g-1}}\frac{1}{\bar{\lambda}^g_i} \nonumber\\
&=& \sum\limits_{i=2}^{2(m+1)^{g-1}} \left(
\frac{1}{\bar{\lambda}^g_i}+
\frac{1}{\bar{\lambda}^g_{i+2(m+1)^{g-1}}}\right)+\frac{1}{\bar{\lambda}^g_{1+2(m+1)^{g-1}}} \nonumber\\
&=&\sum\limits_{i=2}^{2(m+1)^{g-1}}
\frac{\bar{\lambda}^g_i+\bar{\lambda}^g_{i+2(m+1)^{g-1}}
}{\bar{\lambda}^g_i  \bar{\lambda}^g_{i+2(m+1)^{g-1}}
}+\frac{1}{\bar{\lambda}^g_{1+2(m+1)^{g-1}}}.
\end{eqnarray}
Since $\bar{\lambda}^g_i$ and $\bar{\lambda}^g_{i+2(m+1)^{g-1}}$ are
the two roots of Eq.~(\ref{matrix07}), according to the Vieta's
formulas, we have
$\bar{\lambda}^g_i+\bar{\lambda}^g_{i+2(m+1)^{g-1}}=\lambda_{i}^{g-1}+m+1$
and $\bar{\lambda}^g_i
\bar{\lambda}^g_{i+2(m+1)^{g-1}}=\lambda_{i}^{g-1}$. Moreover,
considering $\lambda_{1}^{g}=0$, so
$\lambda^g_{1+2(m+1)^{g-1}}=m+1$. Then, Eq.~(\ref{Hitting12}) is
rewritten as
\begin{eqnarray}\label{Hitting13}
S_g^{(2)}&=& \sum\limits_{i=2}^{2(m+1)^{g-1}}
\frac{\lambda_{i}^{g-1}+m+1}{\lambda_{i}^{g-1}}+\frac{1}{m+1}\nonumber\\
&=&2(m+1)^{g-1}-1+(m+1)\sum\limits_{i=2}^{2(m+1)^{g-1}}\frac{1}{\lambda^{g-1}_i}
+ \frac{1}{m+1}\nonumber\\
&=&2(m+1)^{g-1}-1+(m+1)S_{g-1} + \frac{1}{m+1}\,.
\end{eqnarray}
Using $S_g^{(2)}=S_g-S_g^{(1)}=S_g-2(m-1)(m+1)^{g-1}$ and after some
simplification, Eq.~(\ref{Hitting13}) becomes
\begin{equation}\label{Hitting14}
S_g=(m+1)S_{g-1}+2m(m+1)^{g-1}-\frac{m}{m+1}\,.
\end{equation}
With the initial condition $S_0=\frac{1}{2}$, Eq.~(\ref{Hitting14})
can be solved to yield
\begin{equation}\label{Hitting15}
S_g=\frac{(m+1)^g}{2}+(m+1)^{g-1}(2mg-1)+\frac{1}{m+1}.
\end{equation}
Since $\langle F \rangle_g=2\,S_g$, we have
\begin{equation}\label{Hitting150}
\langle F \rangle_g=(m+1)^g+2(m+1)^{g-1}(2mg-1)+\frac{2}{m+1}.
\end{equation}
We have confirmed this closed-form expression for $\langle F
\rangle_g$ against direct computation from Eqs.~(\ref{Hitting01})
and~(\ref{Hitting03}). For all range of $g$ and different values of
$m$, they completely agree with each other, which shows that the
analytical formula provided by Eq.~(\ref{Hitting150}) is right.
Figure~\ref{Time01} shows the comparison between the numerical and
predicted results, with the latter plotted by the full expression
for the sum in Eq.~(\ref{Hitting150}).

We show next how to represent MFPT, $\langle F \rangle_g$, as a
function of the network order $N_g$, with the aim to obtain the
relation between these two quantities. Recalling $N_g=2\,(m+1)^{g}$,
we have $(m+1)^g=\frac{N_g}{2}$ and
$g=\log_{m+1}\big(\frac{N_g}{2})=\log_{m+1}N_g-\log_{m+1}2$. These
relations enable one to write $\langle F \rangle_g$ in the following
form:
\begin{eqnarray}\label{Hitting16}
\langle F
\rangle_g=\frac{N_g}{2}&+&\frac{2m}{m+1}N_g\log_{m+1}N_g\nonumber\\
&-&\frac{2m\log_{m+1}2+1}{m+1}N_g +\frac{2}{m+1}\,.
\end{eqnarray}

\begin{figure}
\begin{center}
\includegraphics[width=0.7\linewidth,trim=20 8 20 0]{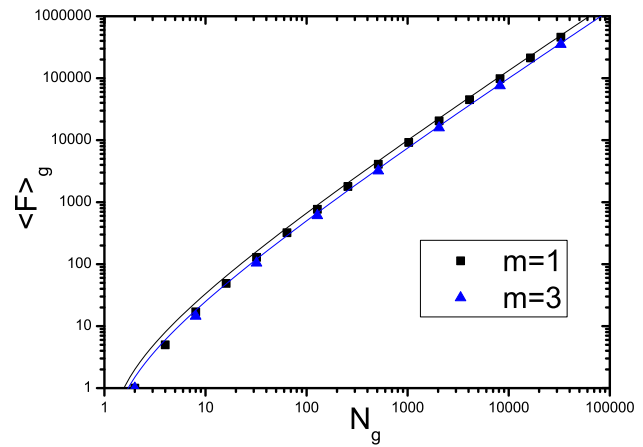}
\end{center}
\caption[kurzform]{\label{Time02} (Color online) Mean first-passage
time $\langle F \rangle_g$ versus the network order $N_g$ on a
log-log scale. The filled symbols describe the analytical results
shown in Eq.~(\ref{Hitting16}). The solid lines represent the
corresponding leading scaling given by Eq.~(\ref{Hitting17}).}
\end{figure}

Equation~(\ref{Hitting16}) unveils the explicit dependence relation
of MFPT on the network order and parameter $m$. For large network,
i.e., $N_g\rightarrow \infty$, we have following expression:
\begin{equation}\label{Hitting17}
\langle F \rangle_g \sim \frac{2m}{(m+1)\ln(m+1)} N_g \ln N_g\,.
\end{equation}
This leading asymptotic $N_g \ln N_g$ dependence of MFPT with the
network order is in contrast to the linear scaling previously
obtained by numerical simulations for scale-free networks, such as
the Apollonian networks~\cite{HuXuWuWa06} and the pseudofractal
scale-free web~\cite{Bobe05}.
Figure~\ref{Time02} shows how the MFPT scales with the network order
for two values of parameter $m$. From Fig.~\ref{Time02}, it is clear
that for properly large network order $N_g$, the dominating term
provided by Eq.~(\ref{Hitting17}) and described by the curve lines
agrees well with the exact formula given by Eq.~(\ref{Hitting16}).

\section{MFPT for trapping in a special network}

In the preceding section, we have shown that the MFPT averaged over
all node pairs, $\langle F \rangle_g$, varies with the network order
$N_g$ as $\langle F \rangle_g \sim N_g \ln N_g$. Below we will show
that the scaling for MFPT averaged over part of node couples may be
different. For this purpose, we will study the trapping issue in a
particular network for $m=1$ case, which is a random-walk problem
where a trap is positioned at a given location. We focus on a
special case with the trap fixed at a node with the largest degree
(hereafter called hub node) absorbing all particles visiting it,
which is a simplistic version of trapping in complex
networks~\cite{Ga04}.

\begin{figure}
\begin{center}
\includegraphics[width=0.5\linewidth,trim=50 0 50 0]{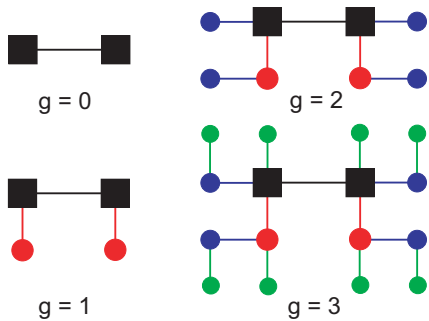}
\caption{(Color online) The growth process for the first three
generations of the particular network corresponding to $m=1$ case.}
\label{RUT1}
\end{center}
\end{figure}

\begin{figure}
\begin{center}
\includegraphics[width=.5\linewidth,trim=10 0 10 0]{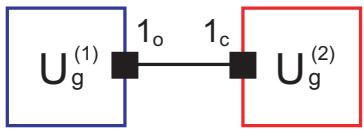}
\caption{(Color online) Second construction method of the network
for $m=1$ case, which highlights self-similarity. The network after
$g+1$ generations, $U_{g+1}$, consists of two replicas of $U_{g}$
denoted separately by $U_g^{(1)}$ and $U_g^{(2)}$, which are
connected to each other by adding a link between two hub nodes with
one in $U_g^{(1)}$ and the other in $U_g^{(2)}$.} \label{Joining}
\end{center}
\end{figure}

For simplicity, we continue to use the notation $U_g$ to represent
the network for $m=1$ case after $g$ iterations. Figure~\ref{RUT1}
illustrates the first several iterations for this network. In fact,
the network has a self-similar structure, which is obvious from the
following equivalent construction method of the network: suppose one
has $U_g$, the next generation of the network, $U_{g+1}$, can be
obtained by joining two $U_g$, see Fig.~\ref{Joining}. We call the
two components, $U_g$, in $U_{g+1}$ the original $U_g$ and duplicate
$U_g$, respectively. For the convenience of description, we label
all node in $U_{g+1}$ using the following way: the nodes in the
original $U_g$ are labeled as $1_{o}$, $2_{o}$, $\ldots$, $N_{g_o}$,
while nodes in the copy of $U_g$ are labeled as $1_{c}$, $2_{c}$,
$\ldots$, $N_{g_c}$. The trap is located at the node belonging to
the original $U_g$ with label $1_{o}$.

Let $T_i(g)$ be the trapping time (TT) of a node $i$ in $U_g$, which
is the expected time for a walker starting from $i$ to first visit
the trap node $1_{o}$. Obviously, for all $g \geq 0$,
$T_{1_{o}}(g)=0$. We first calculate the quantity $T_{1_{c}}(g)$
that is useful for deriving the main result. Since node $1_{c}$ is a
neighbor of the trap node $1_{o}$, according to the result obtained
previously in~\cite{NoKi06}, i.e., Eq. (9) in~\cite{NoKi06}, we have
\begin{equation}\label{Trapping01}
T_{1_{c}}(g)=N_g-1=2^{g+1}-1\,.
\end{equation}

Let $T_{\rm tot}(g)$ denote the sum of trapping time for all nodes
in $U_g$, i.e.,
\begin{equation}\label{Trapping02}
T_{\rm tot}(g)=\sum_{i \in U_g} T_i(g)\,.
\end{equation}
Then, the mean trapping time (MTT), $\langle T \rangle_g$, which is
the average of $T_i(g)$ over all initial nodes distributed uniformly
in $U_g$, is given by
\begin{equation}\label{Trapping03}
 \langle T
\rangle_g=\frac{1}{N_g}\sum_{i \in U_g} T_i(g)=\frac{T_{\rm
tot}(g)}{N_g}\,.
\end{equation}
Thus, to obtain $\langle T \rangle_g$, we should first explicitly
determine the quantity $T_{\rm tot}(g)$, which can be settled using
a recursive way.

According the second construction method of the network, it is not
difficult to express $T_{\rm tot}(g+1)$ in terms of $T_{\rm
tot}(g)$. By definition, we have
\begin{eqnarray}\label{Trapping04}
T_{\rm tot}(g+1)&=&\sum_{i \in U_g^{(1)}} T_i(g+1)+\sum_{i \in
U_g^{(2)}} T_i(g+1)\nonumber \\
&=&\sum_{i \in U_g} T_i(g)+\sum_{i \in U_g^{(2)}}
[F_{i1_c}(g+1)+F_{1_c1_o}(g+1)]\nonumber \\
&=&2\,T_{\rm tot}(g)+N_g\,T_{1_{c}}(g+1)\,.
\end{eqnarray}
Considering $N_g=2^{g+1}$, $T_{1_{c}}(g+1)=2^{g+2}-1$, and the
initial condition $T_{\rm tot}(0)=1$, Eq.~(\ref{Trapping04}) is
inductively to obtain
\begin{equation}\label{Trapping05}
T_{\rm tot}(g)=4^{g+1}-2^{g}(g+3)\,.
\end{equation}
Plugging the last expression into Eq.~(\ref{Trapping03}), we arrive
at the closed-form expression for the MTT on network $U_g$ for the
limiting case of $m=1$,
\begin{equation}\label{Trapping06}
 \langle T
\rangle_g=2^{g+1}-\frac{g+3}{2}\,.
\end{equation}
Clearly, for large network (i.e., $N_g \rightarrow \infty$),
\begin{equation}\label{Trapping06}
\langle T \rangle_g \approx N_g\,,
\end{equation}
implying that the MTT $\langle T \rangle_g$ increases linearly with
the network order.

\begin{figure}
\begin{center}
\includegraphics[width=0.80\linewidth,trim=20 20 20 0]{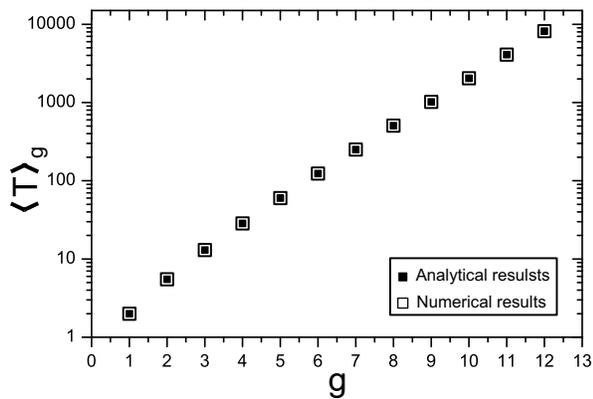}
\end{center}
\caption[kurzform]{\label{MTT01} Mean trapping time $\langle T
\rangle_g$ as a function of the iteration $g$ on a semilogarithmic
scale. The filled symbols are the numerical results obtained by
direct calculation via Eq.~(\ref{Hitting01}), while the empty
symbols correspond to the analytical values provided by
Eq.~(\ref{Trapping05}). The analytical and numerical values are
consistent with each other.}
\end{figure}

We have checked the above analytical result using extensive
simulations. In Fig.~\ref{MTT01}, we plot the simulation results
against Eq.~(\ref{Trapping05}) for different values of $g$. For all
values of $g$, the numerical results are in complete agreement with
the analytical results. Note that the linear dependence of MTT on
network order provided by Eq.~(\ref{Trapping06}) is consistent with
the previously obtained results in~\cite{KiCaHaAr08} by using a
simple approximate method, where it was shown, that for the trapping
problem in scale-free networks having a degree distribution $P(k)
\sim k^{-\gamma}$, when the fixed trap is positioned at a node with
highest degree, the MTT $\langle T \rangle$ varies with the network
order $N$ as $\langle T \rangle \sim N^{\beta}$ with $\beta
=(\gamma-2)/ (\gamma-1)$. Since for an exponential network, such as
the one addressed here, it can be considered as a scale-free network
with $\gamma=\infty$~\cite{AlBa02,DoMe02}, which leads to $\beta=1$,
in agreement with the result given in Eq.~(\ref{Trapping06}). Thus,
the exact linear scaling obtained here confirms the general case,
which was derived based on a simple continuous
approximation~\cite{KiCaHaAr08}.

From the above results, we know that the leading behaviors for
$\langle F \rangle_g$ and $\langle T \rangle_g$ are evidently
different. The former follows $\langle F \rangle_g \sim N_g\,\ln
N_g$, while the latter obeys $\langle F \rangle_g \sim N_g$, less
than that of the former. The distinctness between the two scalings
can be interpreted by the following heuristic arguments on the basis
of the peculiar structure of the network. In the trapping problem
addressed here, the location for trap node is particularly selected,
which lies at a node with largest degree. In fact, the trap node is
the center of the network (see Fig.~\ref{RUT1}): one-half of nodes
(i.e., descendants of the trap node) lies at one side of it,
one-half (i.e., descendants of node $1_c$ including $1_c$ itself) at
the other side. In this case, the walker, irrespective of its
starting point, will visit at most half region of the whole network
before being trapped. On the contrary, for some pairs of nodes, such
as those couples of nodes with both ends being the descendants of
$1_o$ and $1_c$, respectively, the walker may visit large part (even
the entire part) of the network before hitting the target node. This
is the main reason why $\langle T \rangle_g$ is less than $\langle F
\rangle_g$.


Notice that, the random walks discussed in preceding section may be
considered as a trapping problem with the trap uniformly distributed
throughout all nodes on the networks. The different scalings between
$\langle F \rangle_g$ and $\langle T \rangle_g$ can lead us to
conclude that the location of trap has a significant effect on the
leading behavior of the MTT for trapping problem on one particular
network of DURTs, which is in sharp comparison with that of trapping
defined on the $T$ graph notwithstanding its tree structure, where
the MTT is independent of the trap position~\cite{ZhLiZhWuGu09}. The
root of this disparity of the behaviors for random walks on a DURT
and the $T$ graph might lie in their distinct structural properties.
Although they are both trees, the former is small-world with the
average distance behaving logarithmically with its
order~\cite{JuKiKa02}; while the latter is not small-world having an
average distance increasing algebraically with network
order~\cite{ZhLiZhWuGu09}. Particulary, the $T$ graph is a fractal,
while the DURT is not (its fractal dimension is
infinite~\cite{SoHaMa06}). This fractality has also been shown to
distinguish diffusion in scale-free networks. For details, please
see Refs.~\cite{YuKaKi09,ZhXiZhGaGu09}. It should be stressed that
here we only give a possible reason for this difference, the genuine
explanations need further investigation in the future.

\section{Conclusions}

We have studied the standard random walks on a family of
deterministic treelike networks, exhibiting small-world behavior. By
applying the connection between the FPTs and the Laplacian
eigenvalues, we have determined explicitly the MFPT averaged over
all pairs of nodes in the networks. The obtained solution shows that
for large networks with order $N$, the MFPT grows approximatively
with $N$ as $N\ln N$. 
We also presented that compared to the linear scaling of looped
networks, such as the Apollonian networks~\cite{HuXuWuWa06} and the
pseudofractal scale-free web~\cite{Bobe05}, the DURTs studied here
induce a slowing down of diffusion dynamics, providing a useful
insight into random walks on treelike small-world networks.

In the second part of this work, we have investigated the trapping
issue on the a particular network of the DURTs, concentrating on a
special case with the trapping positioned at a node with highest
degree. We have obtained the explicit solution of the ATT, whose
leading behavior varies lineally with network order. Based on the
fact that the standard random walks addressed in the first part of
the work may be looked upon as a general trapping problem with trap
being distributed uniformly on every node in the whole network, we
have drawn a conclusion that the scaling of ATT for trapping depends
on the location of trap. Finally, it is expected that the analytical
computation methods for MFPT and ATT can be extended to other
deterministic media.

\subsection*{Acknowledgment}

We would like to thank Xing Li, Yichao Zhang, and Xiangwei Chu for
their support. This research was supported by the National Natural
Science Foundation of China under Grants No. 60704044, No. 60873040,
and No. 60873070, the National Basic Research Program of China under
Grant No. 2007CB310806, Shanghai Leading Academic Discipline Project
No. B114,the Program for New Century Excellent Talents in University
of China (Grants No. NCET-06-0376), and Shanghai Committee of
Science and Technology (Grants No. 08DZ2271800 and No. 09DZ2272800).
S. Y. G. also acknowledges the support by Fudan's Undergraduate
Research Opportunities Program.

\end{document}